\documentclass[letterpaper,twocolumn,showpacs,prl,aps,superscriptaddress]{revtex4-1}
\usepackage[latin1]{inputenc}
\usepackage{bm}
\usepackage{amsmath}
\usepackage{graphicx}
\usepackage{amssymb}

\begin{document}

\title{Universal local symmetries and non-superposition in classical mechanics}

\author{Ennio Gozzi}
\email{gozzi@ts.infn.it}
\affiliation{Dipartimento di Fisica (sede di Miramare),
Universit\`a di Trieste, Strada Costiera 11, 34014 Trieste, Italy,
\\and \\ Istituto Nazionale di Fisica Nucleare, Sezione di Trieste, Italy.}

\author{Carlo Pagani}
\email{pagani.crl@gmail.it}
\affiliation{Dipartimento di Fisica (sede di Miramare),
Universit\`a di Trieste, Strada Costiera 11, 34014 Trieste, Italy.}

\begin{abstract}
In the Hilbert space formulation of classical mechanics (CM), pioneered by Koopman and von Neumann (KvN), there are potentially more observables that in the standard approach to CM. In this paper we show that actually many of those extra observables are not invariant under a set of universal local symmetries which appear once the KvN is extended to include the evolution of differential forms. Because of their non-invariance, those extra observables have to be removed. This removal makes the superposition of states in KvN, and as a consequence also in CM, impossible. 
\end{abstract}


\maketitle

In classical statistical mechanics the evolution of the probability
densities in phase space $\varrho(q,p,t)$ is given by the Liouville 
equation:
\begin{equation}
\displaystyle i\frac{\partial}{\partial t}\varrho(q,p,t)=\widehat{\cal H}\varrho(q,p,t) \label{Liorho}
\end{equation}
where $\widehat{\cal H}$ is the Liouville operator
\begin{equation}
\widehat{\cal H}= -i\partial_pH(q,p)\partial_q+i\partial_qH(q,p)\partial_p \label{liouv}
\end{equation}
and $H(q,p)$ is the Hamiltonian in phase-space whose coordinates are $(q,p)$. Since the distributions $\varrho(q,p)$
are probability densities, it is sufficient for them to be {\it integrable} functions: $\varrho(q,p)\in L^1$.  
In the 30's Koopman and von Neumann (KvN)~\cite{von Neumann} proposed an alternative approach based on the following four postulates:

\begin{itemize}
\item[{\bf I)}] The system is described by an element $|\psi\rangle$ of an Hilbert space ${\bf H}$.
\smallskip
\item[{\bf II)}] On this Hilbert space the operator $\widehat{q}$ and $\widehat{p}$, associated with the classical variables {\it q} and {\it p}, commutes [$\widehat{q}$,$\widehat{p}$]=0 and can be diagonalized simultaneusly. If we indicate with $\widehat{\varphi}^a$=($\widehat{q}$,$\widehat{p}$) {\it a}=1,2 the diagonalization can be expressed, in a sort of Dirac notation, as:
\begin{equation}
\widehat{\varphi}^a|\varphi^a_{\scriptscriptstyle 0}\rangle=\varphi^a_{\scriptscriptstyle 0}|\varphi^a_{\scriptscriptstyle 0}\rangle \label{autoval}
\end{equation} 
where $\varphi^a_0$ is a particular point in phase space.
\item[{\bf III)}] The evolution of $\psi(q,p)=  \langle \varphi^a|\psi\rangle$ is given by the Liouville equation:
\begin{equation}
\displaystyle i\frac{\partial}{\partial t}\psi(q,p,t)=\widehat{\cal H}\psi(q,p,t).  \label{Liopsi}
\end{equation}
\item[{\bf IV)}] The Liouville probability density $\varrho(q,p)$ is given by $\varrho(q,p,t)=|\psi(q,p,t)|^2$. 
\end{itemize}

As a consequence of the above four postulates we have that also $\varrho(q,p)$ satisfies the Liouville equation.
The KvN formulation resembles very much quantum mechanics (QM) but the operator of evolution, $\widehat{\cal H}$, is not the Schroedinger one, moreover $\widehat{q}$ and $\widehat{p}$ commute, differently than in QM, and the states $\psi(q,p)$ are not the quantum ones.
It was shown in~\cite{gozzi} that the  {\it classical} kernel of propagation $K(\varphi,t|\varphi_i,t_i)$ from an initial phase space configuration $\varphi_i$ at time $t_i$ to a final one $\varphi$ at time $t$, defined as:
\begin{equation}
\displaystyle \psi(\varphi,t)=\int d\varphi_i\, K(\varphi,t|\varphi_i,t_i)\psi(\varphi_i,t_i), \label{evoker}
\end{equation}
can be given the following path integral representation:
\begin{equation}
\displaystyle K(\varphi,t|\varphi_i,t_i)=\int {\cal D}^{\prime\prime}
\varphi^a{\cal D}\lambda_a \,\textrm{exp}\biggl[i\int dt\, {\cal\widetilde L}\biggr] \label{path}
\end{equation}
The symbol ${\cal D}^{\prime\prime}\varphi^a$ indicates the "sum" over all paths in phase space with fixed end points $\varphi$ and $\varphi_i$, while $\lambda_a$ is an auxiliary variable known in statistical mechanics as response field (see for example ref.~\cite{dekke}).
$\cal\widetilde L$ is not the usual lagrangian but an object defined as:
\begin{equation}
{\cal\widetilde L}=\lambda_a\dot{\varphi}^a-{\cal\widetilde H} \label{Lagrangian}
\end{equation}
with the Hamiltonian $\cal\widetilde H$ given by:
\begin{equation}
{\cal\widetilde H}=\lambda_a\omega^{ab}\partial_bH \label{hamilton}
\end{equation}
where $\omega^{ab}$ is the symplectic matrix.
In~\cite{gozzi} the authors did not use the $ \psi(\varphi)$ but the probability densities $\varrho(\varphi)$. Anyhow, as the kernel of propagation of the two, thanks to the postulate {\bf III)}, is the same, the path-integral turns out to be the same for both $ \psi(\varphi)$ and $\varrho(\varphi)$ 
From (\ref{path}), using the time slicing technique, it is easy~\cite{gozzi} to derive the following commutators:
\begin{equation}
[\widehat{\varphi}^a,\widehat{\varphi}^b]=0  \qquad\quad [\widehat{\varphi}^a,\widehat{\lambda}_b]=i\delta_b^a  \qquad\quad [\widehat{\lambda}_a,\widehat{\lambda}_b]=0  \label{comm}
\end{equation}
In order to satisfy the second relation in (\ref{comm}), we can give the following representation for $\widehat{\lambda}_a$, $\widehat{\lambda}_a=-i\frac{\partial}{\partial \varphi^a}$. With this representation we get that the Hamiltonian $\cal\widetilde H$ in (\ref{hamilton}) is turned into the Liouville operator~\cite{gozzi}:
\begin{equation}
 \widetilde{\cal H} \longrightarrow {\widehat{\cal H}}=-i(\partial_p H)\partial_q + i(\partial_q H)\partial_p
\end{equation}  
In the Hilbert space of KvN we can introduce various scalar products~\cite{gozzi-deotto-mauro} but we will stick to the one in which $\widehat\varphi$ and $\widehat\lambda$ are hermitian.
In a more mathematically advanced formulation of hamiltonian mechanics~\cite{mardsen}, it is common to extend the KvN formalism to differential forms on phase-space, i.e.: to objects like $\psi(\varphi,d\varphi)$. The operator which makes the evolution of these forms is called~\cite{mardsen} the {\it Lie derivative of the Hamiltonian flow} and is usually indicated by the simbol ${\cal L}_{(dH)^\sharp}$. It is the sum of two parts, the first one being the Liouvillian (or the Hamiltonian vector fields~\cite{mardsen}) which acts on the zero forms $\psi(\varphi)$ and a second one which is needed for higher forms. Like we did for the Liovillian, it is easy~\cite{gozzi} to give a path-integral construction for the Lie derivative of the Hamiltonian flow. Let us indicate the basis $d\varphi^a$ of the forms with grassmanian variables $c^a$, the path-integral (\ref{path}) is then turned in the following one:
\begin{equation}
 K(\varphi,c,t|\varphi_i,c_i,t_i)= \int {\cal D}^{\prime\prime}\varphi^a {\cal D} \lambda_a {\cal D}^{\prime\prime}c^a {\cal D} \bar{c}_a \exp[i\int dt \widetilde{\cal L}]. \label{cpi1}
\end{equation} 
$\bar{c}_a$ are extra grassmanian variables and $\widetilde{\cal L}$ is:
\begin{equation}
 \widetilde{\cal L} \equiv \lambda_a \dot{\varphi}^a+i\bar{c}_a 
 \dot{c}^a-\widetilde{\cal H}\label{lagro9}
\end{equation}
where:
\begin{equation}
 \widetilde{\cal H} = \lambda_a \omega^{ab} \partial_b H+i\bar{c}_a\omega^{ab} \partial_b\partial_d H c^d. \label{hamiltoniana}
\end{equation}
The commutators which we can derive from (\ref{cpi1}) for $c^a$ and $\bar{c}_a$ are the following ones:
\begin{equation}
[c^a,\bar{c}_b]_{+}=\delta^a_b  \qquad\quad [c^a,c^b]_{+}=0  \qquad\quad [\bar{c}_a,\bar{c}_b]_{+}=0  \label{comm-grass}
\end{equation}
where with $[(\cdot) ,(\cdot) ]_{+}$ we indicate the anticommutators. As we have already said $c^a$ can be identified with the basis $d\varphi^a$ of the forms, in a similar manner $\bar{c}_a$ can be identified with the basis of antisymmetric tensors. The whole set of auxiliary variables, $(\lambda_a,c^a,\bar{c}_a)$, has a geometrical meaning which has been studied in detail in~\cite{regini}. With this geometrical interpretation and the operatorial representation of $\widehat{\varphi}$, $\widehat{c}$, $\widehat{\bar{c}}$ and $\widehat{\lambda}$, the Hamiltonian $\widetilde{\cal H}$ in (\ref{hamiltoniana}) is turned into an operator which coincides with the Lie-derivative ${\cal L}_{(dH)^\sharp}$~\cite{gozzi} of the Hamiltonian flow. The whole Cartan calculus can be reproduced by using the above commutators and seven charges associated to global universal symmetries of the Hamiltonian $\widetilde{\cal H}$. More details can be found in~\cite{gozzi}. Among these global symmetries a particular important role is played by the following two charges:
\begin{eqnarray}
 Q_H \equiv i\widehat{c}^a \widehat{\lambda}_a-\widehat{c}^a \partial_a H  \\
 \bar{Q}_H \equiv i\widehat{\bar{c}}_a \omega^{ab} \widehat{\lambda}_b+\widehat{\bar{c}}_a \omega^{ab} \partial_b H   \label{cariche susy}
\end{eqnarray}
whose anticommutators is:
\begin{equation}
 [Q_H,{\bar Q}_H]_{+}=2i \widehat{\widetilde{\cal H}}.  \label{susy}
\end{equation}
As they close on the Hamiltonian we can genuinely call them supersymmetry charges. Geometrically they are associated to the equivariant cohomology of the Hamiltonian flow~\cite{mardsen} or in simpler terms to the exterior derivative on constant energy surfaces. The exterior derivative on the whole phase space is given by another of the seven charges we mentioned earlier. It has been called BRS charge in analogy with gauge theories and its expression is:
\begin{equation}
 Q^{BRS}=i\widehat{c}^a \widehat{\lambda}_a \label{brs}
\end{equation}
Its symplectic dual~\cite{mardsen} is:
\begin{equation}
 \bar{Q}^{BRS}=i\widehat{\bar{c}}_a \omega^{ab} \widehat{\lambda}_b \label{anti brs}
\end{equation}
which has been named  antiBRS~\cite{gozzi}. Differently than the supersymmetric charges, the BRS and antiBRS charges anticommutes among themselves:
\begin{equation}
 [Q^{BRS},\bar{Q}^{BRS}]_{+}=0.
\end{equation}
We have briefly introduced the BRS and antiBRS charges because they will play a role in the construction we are going to make.

Let us now return to the four postulates {\bf I)}, {\bf II)}, {\bf III)}, {\bf IV)} of the KvN we have listed at the beginning. With respect to QM there is one postulate missing: the one that tells us which are the {\it observables} of the theory.
In this operatorial formulation the most "natural" postulate for this aspect of the theory should be:
\begin{itemize}
\item[{\bf V)}] The observables of the theory are the hermitian operators acting on the KvN Hilbert space.
\end{itemize}
This postulate does not seem to coincide with the standard one of  CM where the observables are taken to be the real functions of $\varphi^a$, i.e.: $O(\varphi^a)$. These, at the operatorial level, would become those hermitian operators which are functions only of $\hat{\varphi}^a$, i.e.: 
\begin{equation}
 O^{\dagger}(\widehat \varphi)=O(\widehat \varphi) \label{operatori-cm}
\end{equation}
According to postulate {\bf V} there are many more observables than just those in 
(\ref{operatori-cm}).Once a scalar product is given  \cite{gozzi-deotto-mauro}, the observables  are  the hermitian combinations  of all operators $\widehat{\varphi}$, $\widehat{c}$, $\widehat{\bar{c}}$ and $\widehat{\lambda}$ present in the KvN approach, i.e.:
\begin{equation}
 O^{\dagger}(\widehat \varphi, \widehat \lambda, \widehat c, \widehat{\bar{c}})=O(\widehat \varphi, \widehat \lambda, \widehat c, \widehat{\bar{c}}), \label{operatori-cpi}
\end{equation}
 Note that they present a feature which should not be present in CM. The feature, even if we restrict to the sector $\widehat{c}=\widehat{\bar{c}}=0$, i.e.:
\begin{equation}
 O^{\dagger}(\widehat \varphi, \widehat \lambda)=O(\widehat \varphi, \widehat \lambda), \label{operatori-kvn}
\end{equation}
is that there are a lot of them which do not commute among themselves since $\widehat \varphi$ and $\widehat \lambda$ do not commute (see (\ref{comm})). This "non-commuting" feature, combined with the superposition principle naturally present in the Hilbert space approach (postulate {\bf I)}), would lead to interference effects which have never been detected in CM.\\
{\it Which is the way out?}. One way could be to replace postulate {\bf V)} with the following one:
\begin{itemize}
\item[{\bf  $\widetilde{\rm {\bf V}}$)}] The observables of the theory are the hermitian operators functions only of ${\widehat {\varphi}}^a$, i.e.:
\begin{equation}
 O^{\dagger}(\widehat \varphi)=O(\widehat \varphi) \label{operatori-cm1}
\end{equation}
\end{itemize}
We do not like to do that because it is like postulating the non-interference effect of CM. What we will prove in this paper is that, using postulate {\bf V)}, together with the other four, we can derive $\widetilde{\rm {\bf V}}{\bf )}$ as a {\it theorem}. The tools which will allow us to do that are some hidden {\it local} symmetries present in the KvN formalism.

Let us first go back to the Lagrangian (\ref{lagro9}) and build the analog of the generating functional:
\begin{equation}
 Z[J]|_{J=0}=\int {\cal D}\varphi^a {\cal D} \lambda_a {\cal D} c^a {\cal D} \bar{c}_a e^{i \int \widetilde{\cal L} dt} \label{generatore}
\end{equation}
Because of the supersymmetry (Susy) (\ref{susy}) present in this formalism, we can use some of the tools developed for Susy like for example the one of superfield~\cite{salam}. This object is something like a multiplet which assembles together all the variables $\varphi^a$, $c^a$, $\bar{c}_a$ and $\lambda_a$. To build it we have first to extend time to two grassmanian partners of it: $\theta$ and $\bar\theta$. The superfield $\Phi$ is a function of $t$, $\theta$ and $\bar\theta$ defined as follows:
\begin{equation}
 \Phi^a(t,\theta,\bar\theta) \equiv \varphi^a+\theta c^a+\bar{\theta} \omega^{ab}\bar{c}_b+i \bar{\theta} \theta \omega^{ab}\bar{\lambda}_b.  \label{superfield}
\end{equation}
Using the superfield is easy to prove that the action associated to the Lagrangian (\ref{cpi1}) can be written as:
\begin{equation}
 \int_{t_0}^{t} \widetilde{\cal L}(\varphi^a,c^a,\bar{c}_a,\lambda_a)= \int_{t_0}^{t} i dt d\theta d\bar{\theta} L[\Phi,\dot{\Phi}]+(s.t.) \label{lagr super}
\end{equation}
where L is the standard Lagrangian of our system associated to the Hamiltonian H which is needed to build the $\widetilde{\cal H}$ of (\ref{hamiltoniana}) and the  $(s.t.)$ is just a surface term. Using (\ref{lagr super}) the generating functional $Z$ in (\ref{generatore}) can be written in a more compact form as:
\begin{equation}
 Z[0]= \int  {\cal D} \Phi^a e^ {i \int i dt d\theta d\bar{\theta} L[\Phi^a,\dot{\Phi}^a]}  \label{super generatore}
\end{equation}
where we have only dropped the surface term. A more detailed presentation of the last steps can be found in~\cite{geom deq}. The formalism presented in (\ref{super generatore}) has the {\it global} invariances  we talked about before: BRS, antiBRS, Susy and three others but it also has some {\it local} symmetries.\\
For example let us consider the following transformation:
\begin{equation}
 \left\lbrace \begin{array}{cc} {\varphi^a} & \longrightarrow \varphi^a+\varepsilon(t) \theta c^a \\  {c^a}  & \longrightarrow  {c^a}-\varepsilon(t) c^a \end{array} \right.  \label{trasf1}
\end{equation}
where $\varepsilon(t)$ is an infinitesimal parameter depending on $t$ and $\theta$ is the grassmanian partner of time used in the superfield (\ref{superfield}). It is easy to check that the superfield (\ref{superfield}) remains invariant under (\ref{trasf1}) and the same for its time derivative. There are many other similar invariances like these. In this paper we will present only two more besides (\ref{trasf1}). They are:
\begin{equation}
 \left\lbrace \begin{array}{cc} {\varphi^a} &\longrightarrow \varphi^a+\varepsilon(t) \bar{\theta} \omega^{ab} \bar{c}_b \\  {\bar{c}_b} & \longrightarrow {\bar{c}_b}-\varepsilon(t) \bar{c}_b \end{array} \right.  \label{trasf2}
\end{equation}
and
\begin{equation}
 \left\lbrace \begin{array}{cc} {\varphi^a} & \longrightarrow \varphi^a+i \varepsilon(t) \bar{\theta} \theta \varphi^a \\  {\lambda_b} & \longrightarrow {\lambda_b}-\varepsilon(t) \omega_{bc} \varphi^c \end{array} \right.  \label{trasf3}
\end{equation} 
It is easy to prove that the only object invariant under all three of them is the superfield and its time derivative. Actually the first transformation (\ref{trasf1}) leaves invariant also the sub-piece $:\varphi^a+\theta c^a$ but this is not invariant under (\ref{trasf2}) and (\ref{trasf3}). Similarly (\ref{trasf2}) leaves invariant, besides the superfield, also the sub-piece: $\varphi^a+\bar{\theta} \omega^{ab} \bar{c}_b$, but this is not left invariant by (\ref{trasf1}) and (\ref{trasf3}). Similarly for (\ref{trasf3}). So the only object that is simultaneously invariant under (\ref{trasf1}), (\ref{trasf2}) and (\ref{trasf3}) is the superfield and its time derivative. As a consequence also the action in (\ref{super generatore}) is invariant and so we can say that (\ref{trasf1}), (\ref{trasf2}), (\ref{trasf3}) are local symmetries of our system.\\
Let us now turn to the observables. As there are local symmetries in our system, the acceptable observables are only those invariant under the same local symmetries. As the superfield and its time derivative are the only objects invariant under (\ref{trasf1}), (\ref{trasf2}), (\ref{trasf3}) we have that the only acceptable observables are:
\begin{equation}
 \widetilde{O}(\widehat{\Phi}^a, {\dot{\widehat{\Phi}}^a})
\end{equation}
Using the equation of motion for $\Phi^a$, \cite{gozzi} i.e.: \break $\dot{\Phi}^a=\omega^{ab} \partial_b H[\Phi]$ we can replace $\dot{\Phi}$ by a function of $\Phi$ in $\widetilde O$, and so we can conclude that the acceptable observables are of the form:
\begin{equation}
O(\widehat{\Phi}^a)  \label{osservabili-h}
\end{equation}
with $O \neq \widetilde O$. At this point the reader may point out that the $O(\widehat{\Phi}^a)$ are not the $O(\widehat{\varphi}^a)$ which were the observables we wanted to get. Actually the (\ref{osservabili-h}) are observables which depend explicitly on the two grassmanian partners of time $\theta$ and $\bar \theta$. So with respect to these extra times we can say that the observables (\ref{osservabili-h}) are in the "Heisenberg" picture. The "Hamiltonians" associated to the extra-time variables $\theta$ and $\bar \theta$ are the $Q^{BRS}$ and $\bar{Q}^{BRS}$ of (\ref{brs}) and (\ref{anti brs}) because they generate a translation in $\theta$ and $\bar \theta$ respectively~\cite{gozzi}. So the transformation from the Heisenberg picture to the Schroedinger one in $\theta$ and $\bar \theta$, is:
\begin{equation}
 e^{-\theta Q^{BRS}-\bar{Q}^{BRS} \bar{\theta}} O_H(\widehat{\Phi}^a)  e^{\theta Q^{BRS}+ \bar{Q}^{BRS} \bar{\theta}}  \equiv
 O_S=O(\widehat{\varphi})  \label{cambio pittura}
\end{equation}
The last step in eq. (\ref{cambio pittura}) was proved in~\cite{geom deq}. This is the most important step of the theorem we wanted to prove. Basically (\ref{cambio pittura}) tells us that the acceptable observables are the $O(\widehat{\Phi}^a)$ and these are nothing else than the Heisenberg picture in $\theta$ and $\bar \theta$ of the standard observables of CM $O(\widehat{\varphi}^a)$.

Let us note that in the picture above there is an operator, $\widehat{\varphi}^a$, which commutes with all the observables and which is not a multiple of the identity. This triggers the mechanism known as {\it superselection} (for a review see~\cite{fonda ghirardi} and~\cite{roman}). It says that the physical Hilbert space of the system is given by an eigenvariety of the superselection operator, i.e.: $\widehat{\varphi}^a$. In our case we have:
\begin{equation}
 \widehat{\varphi}^a |\varphi^a_0 \rangle= \varphi^a_0 |\varphi^a_0 \rangle \label{eigen1}
\end{equation}
where $\varphi^a_0$ is a particular point in the phase space. This is the whole Hilbert space, basically a Dirac delta state:
\begin{equation}
 \langle \varphi^a|\varphi^a_0 \rangle= \delta(\varphi^a-\varphi^a_0) \label{dirac state}
\end{equation}
Another Hilbert space shall be given by another eigenvariety:
\begin{equation}
 \widehat{\varphi}^a |\varphi^a_1 \rangle= \varphi^a_1 |\varphi^a_1 \rangle  \label{eigen2}
\end{equation}
where $\varphi^a_1$ is another  particular point in the phase space. Of course, as (\ref{eigen1}) and (\ref{eigen2}) are different Hilbert spaces, we cannot do linear superposition among them , i.e.: the state
\begin{equation}
 |\widetilde{\varphi^a} \rangle \equiv |\varphi^a_0 \rangle+|\varphi^a_1 \rangle
\end{equation}
is not a {\it physical} one. This is the basic reason why in CM there is no superposition and as a consequence no interference. This analysis is a standard one but it should be done in an Hilbert space framework like the KvN.\\
The careful reader could point out that we should have moved in the Schoeredinger picture of $\theta$,$\bar \theta$ also for the states and not just for the observables. The transformation is:
\begin{eqnarray}
 \widetilde{\psi}_S(t,\theta,\bar\theta)  & = & e^{\theta Q^{BRS} +\bar{Q}^{BRS} \bar\theta}\psi (\varphi,c,t)= \nonumber \\
 &=& {\psi}(\varphi^a+\theta c^a+\bar{\theta}\omega^{ab}\bar{c}_b,c)  \label{psi sh}
\end{eqnarray}
The same careful reader could point out that also the operator $\widehat{c}^a$ commutes with all observables and so it is also a superselection operator that has to be diagonalized:
\begin{equation}
 \widehat{c}^a |c^a_0 \rangle= c^a_0 |c^a_0 \rangle.
\end{equation}
As a consequence the Hilbert space is made of the state:
\begin{equation}
 \widetilde{\psi}=\delta(\varphi-\varphi_0) \delta(c-c_0) \label{eigenstate}
\end{equation}
Is this state of the form (\ref{psi sh})? The only manner to achieve that is to put $c=\bar{c}=0$  and so the (\ref{eigenstate}) is reduced to:
\begin{equation}
 \widetilde{\psi}=\delta(\varphi-\varphi_0) \delta(c) \label{eigenstate1}
\end{equation}
The $\delta(c)$ restricts the states to the zero-form states and so $\widetilde{\psi}$ is isomorphic to :
\begin{equation}
{\psi}=\delta(\varphi-\varphi_0). \label{eigenstate2}
\end{equation}
and these are the typical states of CM (\ref{dirac state}).

{\it Conclusion.} The main conclusion we like to draw is that the non-superposition and non-interference in CM are basically due to some local universal invariances ((\ref{trasf1}), (\ref{trasf2}), (\ref{trasf3})) present in CM. Let us point out that in QM we do not have anymore those invariances because, as it has been proved in~\cite{geom deq}, QM is obtained from CM by a dimensional reduction which sends $(\theta,\bar\theta) \longrightarrow 0$. If these variables, $\theta$ $\bar\theta$, are zero the local symmetries disappear and the transformations (\ref{trasf1}), (\ref{trasf2}), (\ref{trasf3}) are reduced to the identity. Somehow we could say that the disappearance of these symmetries seems to trigger the typical interference effects of QM. To shed more light on this we would like in the future to understand the {\it geometry} behind these local symmetries like we did in the past for the global ones~\cite{gozzi, regini} present in $\tilde{\cal L}$.	 

{\it Acknowledgements.} We would like to thank G.C. Ghirardi, S. Kuzenko, M. Reuter and E. Spalucci for helpful discussions. This work has been supported by grants from MIUR and INFN. E.G. likes to dedicate this work to the memory of Silviu at Pepinu d'Agosta at Cugos.

\def\polhk#1{\setbox0=\hbox{#1}{\ooalign{\hidewidth
  \lower1.5ex\hbox{`}\hidewidth\crcr\unhbox0}}} \def\cprime{$'$}
  \def\cprime{$'$}

\end{document}